\documentclass[journal=nalefd, manuscript=letter, layout=twocolumn]{achemso}

\usepackage{amssymb, physics, titlesec}
\usepackage[separate-uncertainty = true]{siunitx}
\usepackage[hidelinks]{hyperref}
\usepackage{xcolor}
\usepackage[normalem]{ulem}

\makeatletter
\let\l@addto@macro\relax
\makeatother
\usepackage[fontsize=10pt]{scrextend}
\mciteErrorOnUnknownfalse

\titleformat{\subsection}[runin]{\bfseries}{}{}{}[]
\let\oldmaketitle\maketitle
\let\maketitle\relax

\newcommand{\supp}[1]{Supporting Section~S#1}

\title{Heralded spectroscopy reveals exciton-exciton correlations in single colloidal quantum dots}

\author{Gur Lubin}
\affiliation{Deptartment of Physics of Complex Systems, Weizmann Institute of Science, Rehovot, Israel}
\altaffiliation{Contributed equally to this work}

\author{Ron Tenne}
\affiliation{Deptartment of Physics of Complex Systems, Weizmann Institute of Science, Rehovot, Israel}
\alsoaffiliation{Department of Physics and Center for Applied Photonics, University of Konstanz, Konstanz, Germany}
\altaffiliation{Contributed equally to this work}

\author{Arin Can Ulku}
\author{Ivan Michel Antolovic}
\author{Samuel Burri}
\affiliation{School of Engineering, École polytechnique fédérale de Lausanne (EPFL), Neuchâtel, Switzerland}

\author{Sean Karg}
\author{Venkata Jayasurya Yallapragada}
\affiliation{Deptartment of Physics of Complex Systems, Weizmann Institute of Science, Rehovot, Israel}

\author{Claudio Bruschini}
\author{Edoardo Charbon}
\affiliation{School of Engineering, École polytechnique fédérale de Lausanne (EPFL), Neuchâtel, Switzerland}
\email{edoardo.charbon@epfl.ch}

\author{Dan Oron}
\affiliation{Deptartment of Physics of Complex Systems, Weizmann Institute of Science, Rehovot, Israel}
\email{dan.oron@weizmann.ac.il}

\begin{document}
\twocolumn[
\begin{@twocolumnfalse}
\oldmaketitle
\begin{abstract}
Multiply-excited states in semiconductor quantum dots feature intriguing physics and play a crucial role in nanocrystal-based technologies. While photoluminescence provides a natural probe to investigate these states, room temperature single-particle spectroscopy of their emission has so far proved elusive due to the temporal and spectral overlap with emission from the singly-excited and charged states. Here we introduce biexciton heralded spectroscopy, enabled by a single-photon avalanche diode array based spectrometer. This allows us to directly observe biexciton-exciton emission cascades and measure the biexciton binding energy of single quantum dots at room temperature, even though it is well below the scale of thermal broadening and spectral diffusion. Furthermore, we uncover correlations hitherto masked in ensembles, of the biexciton binding energy with both charge-carrier confinement and fluctuations of the local electrostatic potential. Heralded spectroscopy has the potential of greatly extending our understanding of charge-carrier dynamics in multielectron systems and of parallelization of quantum optics protocols.
\end{abstract}
\end{@twocolumnfalse}
]

% -----------------------------------------------------------------------------
% --------------------------------- Figures -----------------------------------
% -----------------------------------------------------------------------------
\begin{figure}[t]
    \centering
    \includegraphics[width=\linewidth]{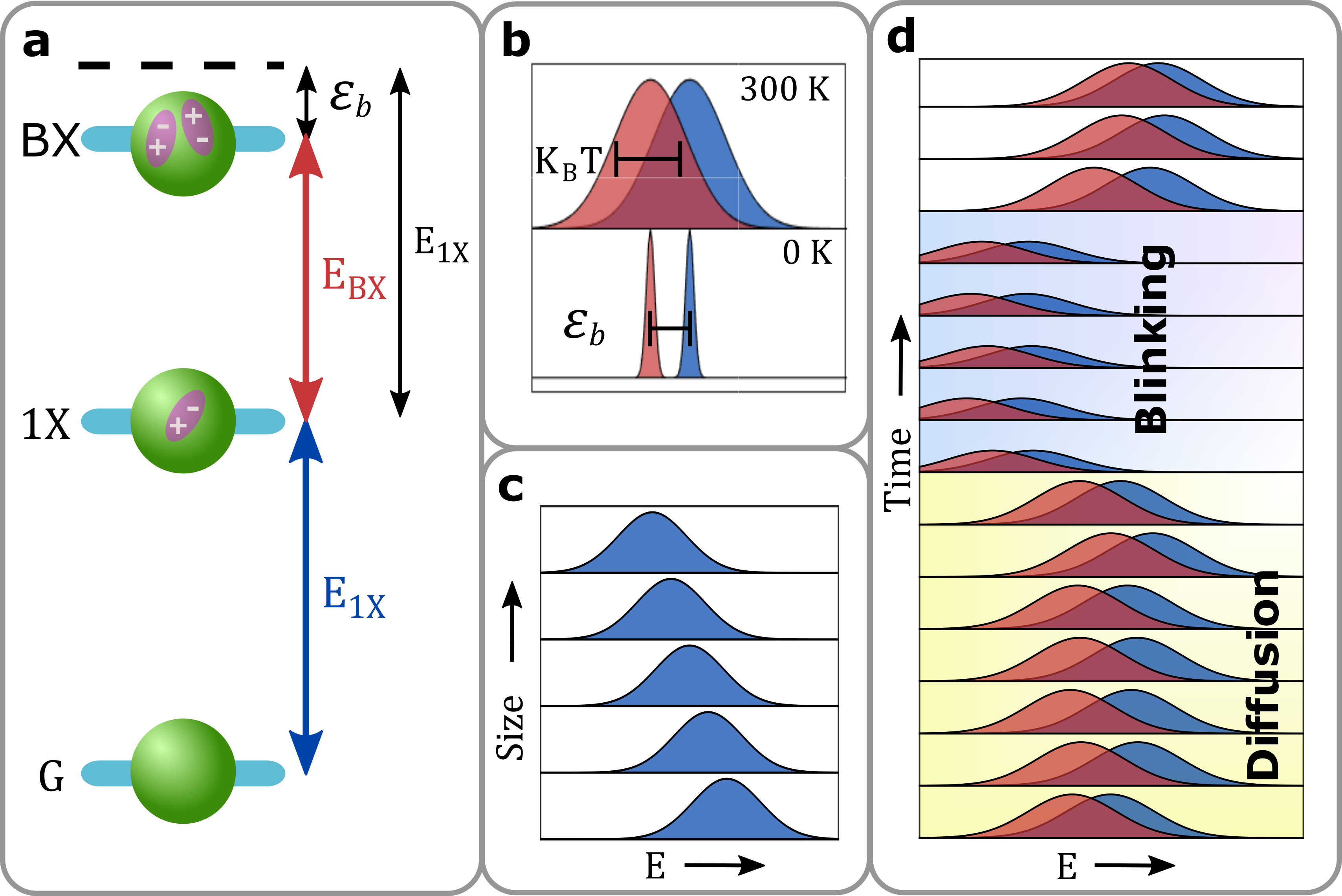}
    \caption{\textbf{Obstacles in measuring the 1X-BX energy ladder. a)} The energy diagram of the ground, single exciton (1X) and biexciton (BX) states in a nanoparticle. The energy difference between the BX and 1X states ($E_{BX}$) is smaller than the difference between the 1X and ground states ($E_{1X}$) by the biexciton binding energy ($\varepsilon_b$). \textbf{b)} Schematic of the thermal broadening of spectral lines (intensity normalized for clarity). At room temperature the 1X-ground (blue) and BX-1X (red) transitions emission lines are broadened to approximately ${\sim}k_{B}T \approx \SI{26}{mev} > \varepsilon_b$. As a result, the two emission spectra substantially overlap.
    \textbf{c)} Schematic dependence of the 1X energy on nanoparticle size. An ensemble measurement (with a $\pm5\%$ size variance) at room temperature roughly includes a mixture of the depicted spectra. \textbf{d)} Scheme of spectral drift in the emission lines of a single nanocrystal throughout the measurement time. Apart from the stochastic random walk of the spectral lines (diffusion), discrete spectral jumps typically accompany blinking events.}    
    \label{fig:intro}
\end{figure}

\begin{figure*}[t]
    \centering
    \includegraphics[width=.9\linewidth]{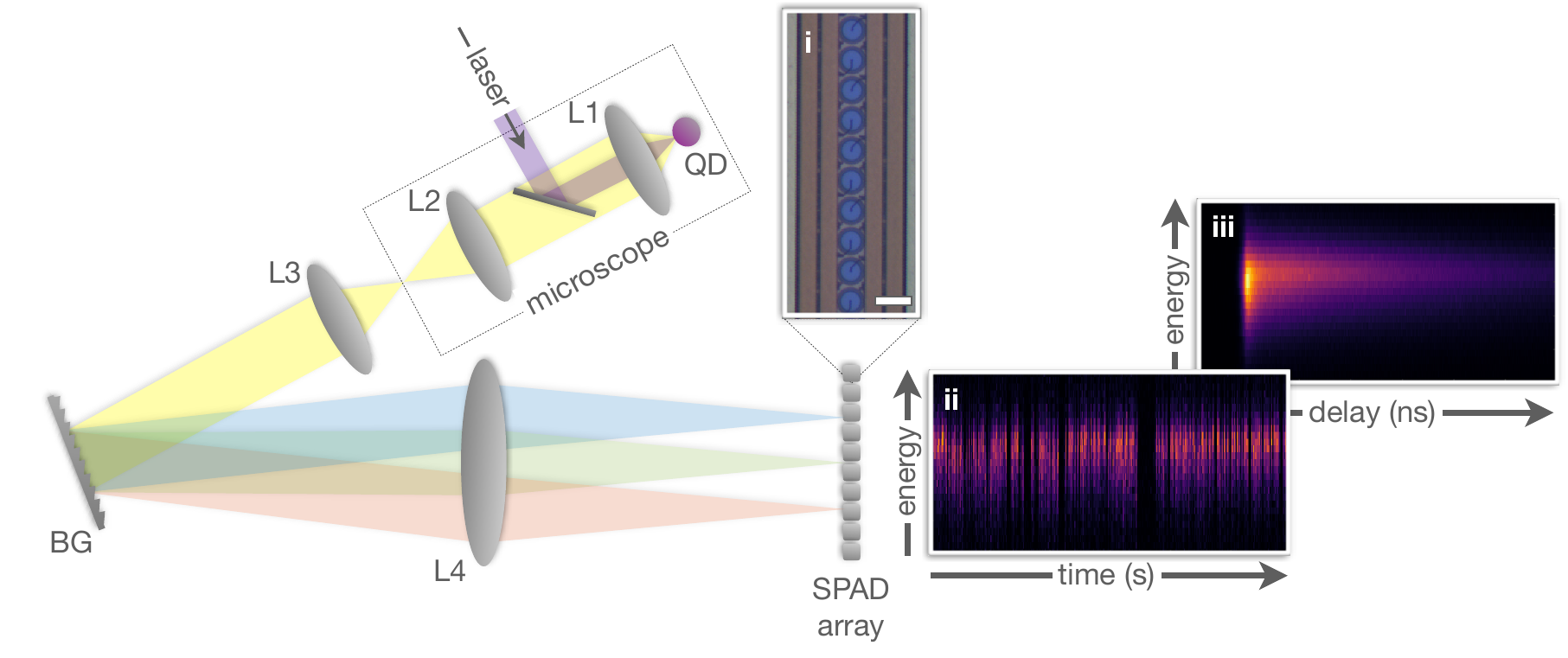}
    \caption{\textbf{Sketch of the spectroSPAD experimental apparatus.} A microscope objective is used to focus a pulsed laser beam on a single QD and collect the emitted fluorescence. At the microscope output, light is passed through a spectrometer setup with a linear SPAD array detector. Inset (i) is an optical image of part of the detector array. Each blue circle represents a single pixel. Scale bar is \SI{30}{\mu m}. Insets (ii) and (iii) show two possible analyses of a single nanocrystal spectroSPAD measurement. In (ii), detections are binned according to detection time in the millisecond scale (horizontal) and photon energy (vertical); whereas in (iii) the horizontal axis represents the detection temporal delay from the preceding excitation laser pulse in nanosecond scale. Color-scale corresponds to the number of counts in each temporal-energy bin. Optical elements: objective (L1), tube (L2), collimating (L3) and imaging (L4) lenses; blazed grating (BG).}
    \label{fig:system}
\end{figure*}

\begin{figure*}[t]
    \centering
    \includegraphics[width=.9\linewidth]{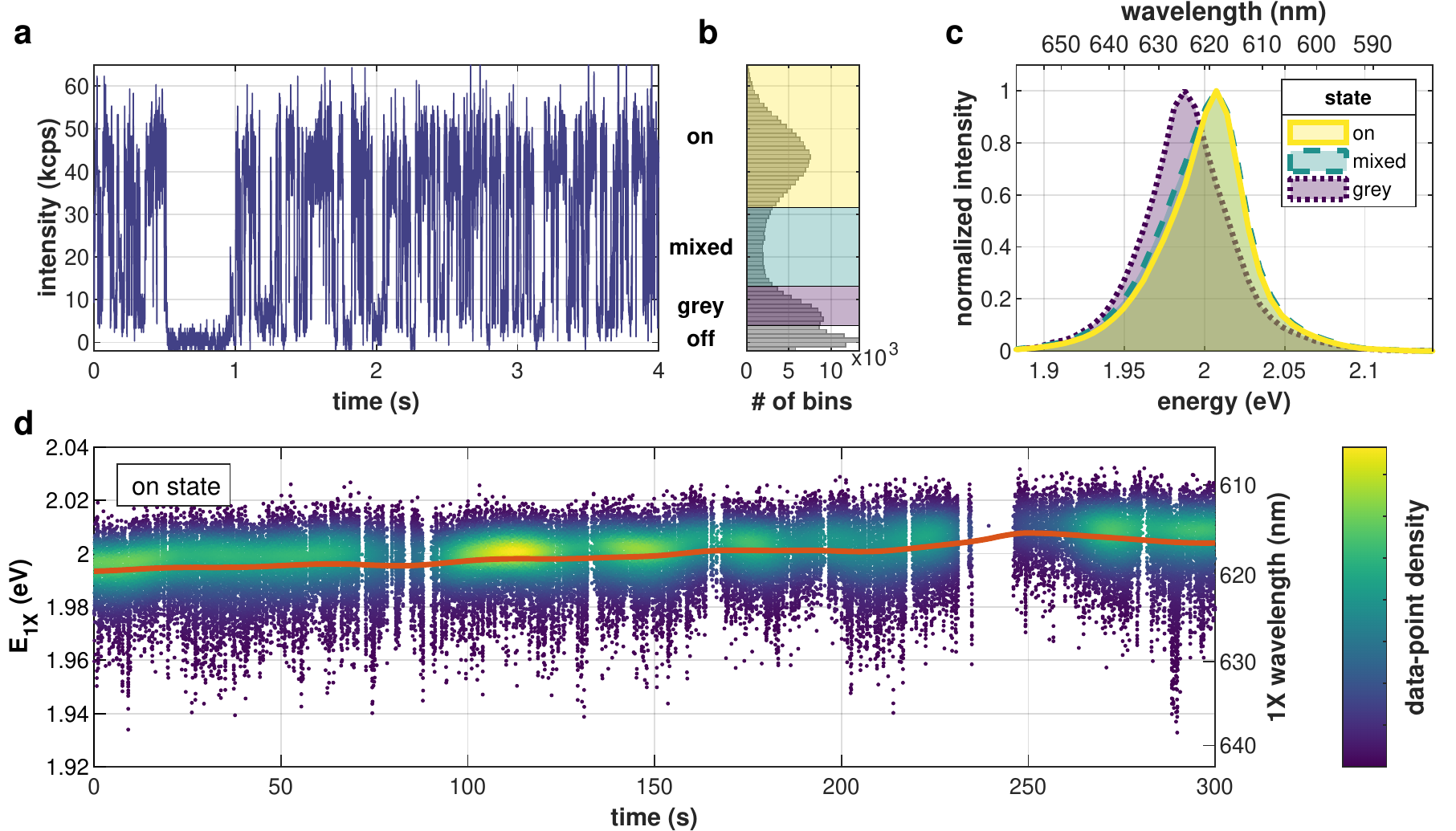}
    \caption{\textbf{Spectral dynamics of a single QD. a)} Total detected fluorescence intensity, collected over all pixels, vs.\ time in a 4-second period (\SI{1}{ms} time-bins). \textbf{b)} Histogram of intensity values over a 5-minute measurement. Intensity states are marked by colored shading: `off', `grey', `mixed' and `on'. \textbf{c)} Spectrum according to intensity gating. Note the `grey' state's red-shift with respect to the `on' state. \textbf{d)} `On' state spectral peak evolution over time. Each point is the momentary mean photon energy for a \SI{1}{ms} time-bin of the `on' state ($\ev{E}_{\SI{1}{ms}}$), colored according to the local density of data-points for clarity. The red line, $\ev{E}_{\SI{10}{s}}$, presents a moving Gaussian-weighted average ($\sigma=\SI{10}{s}$).}
    \label{fig:specDyn}
\end{figure*}

\begin{figure*}[t]
    \centering
    \includegraphics[width=.9\linewidth]{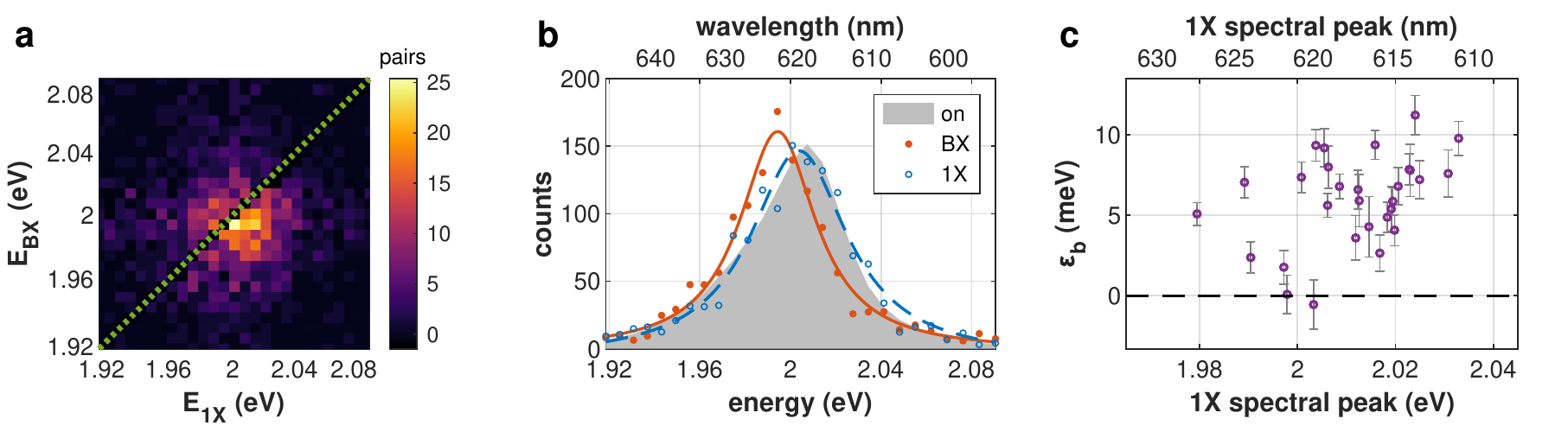}
    \caption{\textbf{Heralded spectroscopy. a)} 2D histogram of photon pairs following the same excitation pulse, according to the energy of the first ($E_{BX}$) and second ($E_{1X}$) photons (vertical and horizontal axes, respectively), over a 5-minute measurement. Dashed green line serves as a guide to the eye, marking same energy for both photons (undetectable by the system) \textbf{b)} Spectrum of the BX (red dots), 1X (blue rings), and all `on' state detections (grey area, normalized). Solid red and dashed blue lines are fits of the BX and 1X spectra, respectively, to a Cauchy-Lorentz distribution. Binding energy, estimated as the difference between BX and 1X spectral peaks, is $\varepsilon_b=\num{9.3+-1.0}~\si{meV}$. \textbf{c)} Binding energy as a function of 1X spectral peak for 30 QDs. Error bars depict 68\% confidence intervals.}
    \label{fig:bx}
\end{figure*}

\begin{figure}[t]
    \centering
    \includegraphics[width=\linewidth]{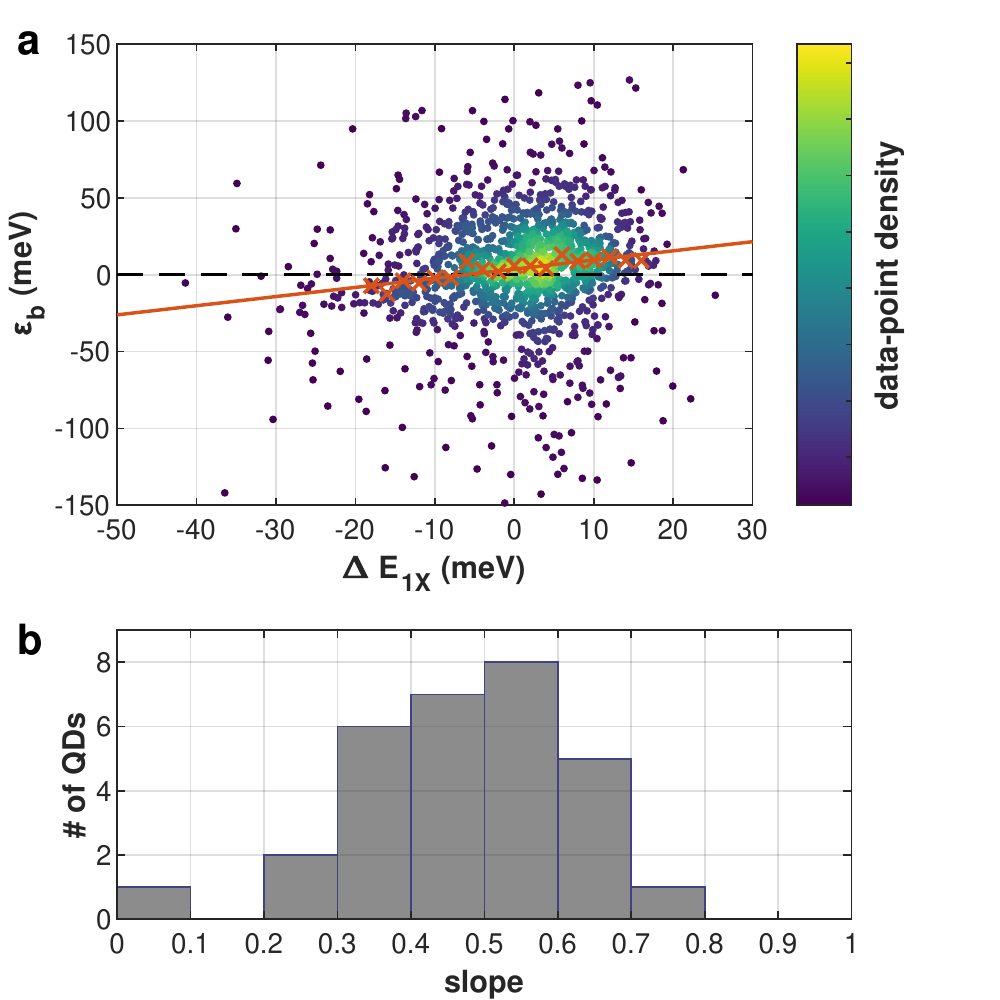}
    \caption{\textbf{BX binding energy fluctuations are correlated with 1X spectral diffusion. a)} Estimated binding energy ($\varepsilon_b$) as a function of the momentary 1X spectral shift ($\Delta E_{1X}$) for a single QD. Each point represents one post-selected BX event, colored according to the local density of data-points for clarity. As a guide to the eye, red crosses mark median binding energy for each \SI{2}{meV} window of $\Delta E_{1X}$. The red line represents a linear fit to these medians (slope $0.59\pm0.08$). \textbf{b)} Histogram of the $\varepsilon_b$ median slope (red line in (a)) for 30 QDs.}
    \label{fig:bx_bySpec}
\end{figure}

% -----------------------------------------------------------------------------
% ------------------------------- main text -----------------------------------
% -----------------------------------------------------------------------------

\section*{Introduction}
Over the past three decades, numerous types of semiconductor nanocrystals with varying compositions, shapes, sizes and structures have been fabricated and studied\cite{Murray2000,Kagan2016,Kovalenko2017}, with some even making their way into mass produced consumer products\cite{Jang2010}. Since the energy of a charge carrier in a nano-confined solid is quantized, nanocrystals are often referred to as `artificial atoms'. In a further analogy to atomic physics, photo-exciting such a quantum dot (QD) generates a Hydrogen-like electron-hole state, an exciton, which is typically bound even at room temperature, due to the increased Coulomb interaction. 
However, unlike atoms and molecules, semiconductor nanocrystals include another readily-excited manifold of states, multiexcitons; multiple electron-hole pair states\cite{Klimov2007}. In the lowest energy multiexcitonic state, the biexciton (BX), a strong exciton-exciton interaction is imposed by the confining potential of the nanoparticle. In the resulting energy ladder of ground, single exciton (1X) and BX states, shown in \autoref{fig:intro}a, the BX state energy is somewhat offset from twice the energy of the 1X state, by the BX binding energy ($\varepsilon_b$)\cite{Klimov2007}. This binding energy is considered positive (attractive interaction) when $E_{BX}<E_{1X}$, where $E_k$ is the energy difference between state $k$ and the state right beneath it in the ladder. 

Cascaded relaxation from the top to the bottom of this ladder can yield a pair of photons; the first around $E_{BX}$ and the second around $E_{1X}$. 
Such an emission from self-assembled Indium Arsenide (InAs) QDs, for example, is a leading candidate for efficiently generating on-demand entangled photon pairs \cite{Senellart2017}. 
On the other hand, avoiding the excitation of the BX state is key for using the same type of QDs as high-purity single-photon sources\cite{Senellart2017}.
More conventional light-based applications that stand to vastly benefit from the incorporation of colloidally synthesized QDs, such as light emitting diodes (LEDs)\cite{Colvin1994,Shirasaki2013}, lasers\cite{Klimov2000,Pelton2018}, displays\cite{Jang2010} and photovoltaics\cite{Kramer2011}, also require characterization and control of the energy and dynamics of the BX state. For instance, non-radiative Auger recombination often dominates the BX relaxation dynamics\cite{Nair2011}. Therefore, to achieve low threshold lasing from QDs, sophisticated heterostructures have been designed to either increase the BX energy via Coulomb repulsion (avoiding its occupation)\cite{Nanda2007}, or to reduce the Auger recombination rate\cite{Todescato2012}. Furthermore, the Auger induced low BX quantum yield sets a saturation boundary for the achievable light fluence of nanocrystal-based LEDs\cite{Wu2019}. 

The above-mentioned interest in the BX-1X ladder inspired substantial spectroscopic efforts to characterize the BX binding energy in multiple material systems such as III-V\cite{Dekel2000,Stevenson2006,Akopian2006}, II-VI\cite{Achermann2003,Oron2006} and lead halide perovskite\cite{Castaneda2016,Shulenberger2019} semiconductor nanocrystals as well as atomically thin films of transition metal dichalcogenides (TMDC)\cite{You2015}. However, conditions under which the BX-1X ladder can be directly probed are restrictive. While at cryogenic temperatures the very stable and narrow 1X and BX emission lines of a single self-assembled InAs or colloidal cadmium chalcogenide QDs can be discerned\cite{Stevenson2006,Akopian2006,Osovsky2009}, in most other cases this was not achieved due to several fundamental limitations. First, at room temperature, both emission lines are thermally broadened well-beyond typical BX binding energies (\autoref{fig:intro}b). Spectral features are further broadened in ensemble measurements due to nanoparticle inhomogeneity  (\autoref{fig:intro}c). Additional inhomogeneous broadening is caused by spectral fluctuations (\autoref{fig:intro}d). In many nanocrystals, this includes not only the spectral diffusion due to spurious electric fields, but also spectral jumps due to other states contributing to fluorescence intermittency\cite{Beyler2013, Antolinez2019}.

More indirect methods to probe the BX state rely on power-dependent measurement of photoluminescence (PL) (either in a time-resolved or quasi-continuous-wave manner)\cite{Oron2006, Sitt2009, You2015, Shulenberger2019} or of transient absorption\cite{Castaneda2016, Steinhoff2018}. While careful modeling and analysis of these measurements provided important spectroscopic information, it has often led to large variance in BX binding energies measured in different studies. For example, while some works found very high (${\sim}\SI{100}{meV}$) BX binding energies in CsPbX\textsubscript{3} (X = Cl, Br, I) nanocrystals\cite{Castaneda2016}, recently a substantially more stringent bound on its magnitude ($\abs{\varepsilon_b}<\SI{20}{meV}$) has been reported\cite{Shulenberger2019,Ashner2019}.
Such discrepancies are mainly due to the difficulty in modeling the different mechanisms that affect PL and absorption at high excitation powers, such as charging, oxidation, blinking and photo-induced damage\cite{Shulenberger2019}. Furthermore, these methods demand disentangling the spectral contributions of the BX and 1X states, which becomes increasingly difficult to perform when these features strongly overlap for small BX binding energies.

While isolating the BX state in the spectral domain alone is a convoluted task, isolating it in the time domain is conceptually simple. A detection of a photon pair emitted from a single nanocrystal (following a short excitation pulse) pinpoints a relaxation cascade\cite{Stevenson2006, Akopian2006}: first from the BX to the 1X state and then from the 1X to the ground state (\autoref{fig:intro}a). Separating the two consecutive emissions according to their detection times, on a nanosecond scale, and measuring their respective spectrum, facilitates a direct measurement of $E_{BX}$ and $E_{1X}$ for a single nanoparticle.
However, currently available instrumentation does not enable practical implementation of this simple scheme. Namely, a standard spectrometer cannot provide the required temporal resolution since it relies on cameras with a maximal frame rate of ${\sim}10^3~\si{fps}$. This can be addressed by replacing the camera with a monolithic single-photon avalanche diode (SPAD) array; a technology that has achieved a considerable performance boost over the past decade\cite{Bruschini2019, Morimoto2020}. In such a novel spectrometer, termed here spectroSPAD, the spectral information of single photons can be measured and correlated with sub-nanosecond temporal resolution.

Here we present the spectroSPAD system and use it to measure spectral correlations in BX-1X emission of single CdSe/CdS/ZnS core/shell/shell QDs with meV precision. Thanks to the temporal resolution and high sensitivity of the spectroSPAD, our measurement scheme overcomes all of the aforementioned obstacles (thermal broadening, spectral diffusion, blinking and low BX yield) and easily separates the BX emission from the misleadingly similar trion (`grey') charged state emission. While for the particular sample under study the average value of the binding energy is ${\sim}\SI{6}{meV}$, we disentangle the inhomogeneous size effect and show that its value in individual QDs correlates with the 1X band edge transition energy. Furthermore, we follow the temporal fluctuations of the BX binding energy for a single nanocrystal and find that those correlate with the spectral diffusion of the 1X transition.

\section*{Results and Discussion}

\subsection*{Apparatus.}
In recent years, considerable efforts were invested in the design of time-resolved light spectrometers with high sensitivity\cite{Blacksberg2011,Gudkov2015,Krstajic2015,Cheng2019,Hartmann2020}. As a replacement for the standard CCD camera, different research groups adopted photo-multiplier tube (PMT) arrays\cite{Gudkov2015}, superconducting nanowire single photon detectors (SNSPDs)\cite{Kahl2017,Cheng2019,Hartmann2020} or SPAD arrays\cite{Nissinen2011,Blacksberg2011,Krstajic2015,Ghezzi2021}. While these implementations harbor great potential for applications such as Raman spectroscopy and on-chip quantum communications, none is able to provide the combination of high overall detection efficiency, low dark counts, and parallel time and spectrum detection at single-photon level. The spectroSPAD spectrometer (\autoref{fig:system}) achieves precisely that by employing a high-performance linear SPAD array as a detector in a Czerny-Turner spectrometer. While a detailed description of the experimental setup is given in \supp{1}, we provide here a brief account. A microscope with a high numerical aperture objective is used to focus pulsed laser illumination on a single QD, and to collect epi-detected fluorescence. This signal is spectrally filtered from the excitation laser with a dichroic mirror and a dielectric filter (not shown), and imaged by a second lens. This image serves as the input for a spectrometer setup - a 4f system with a blazed grating at the Fourier plane. At the output image plane of the spectrometer, a monolithic linear SPAD array is placed, such that each pixel is aligned with the image of a different wavelength range. The detector's photon detection efficiency (PDE) is ${\sim}8\%$ at \SI{620}{nm} and ${\sim}11\%$ at \SI{530}{nm} (this can be improved, see Discussion) and the median dark count rate (DCR) is ${\sim}33$ counts per second per pixel. We analyze the signal of only 40 out of the 512 pixels available in the array, thereby spanning approximately \SI{80}{nm} around a center wavelength of \SI{620}{nm}. This results in a spectral resolution of ${\sim}\SI{2}{nm}$ ($6{-}\SI{7}{meV}$). Single-photon detections are time-tagged by an array of 64 time-to-digital converters (TDCs) programmed in the firmware of a field-programmable gate-array (FPGA). The TDCs are synchronized with the excitation laser. Finally, time and wavelength tagged data is analyzed with a dedicated MATLAB script.

Insets (ii) and (iii) show possible visualizations of fluorescence data collected by the system from a single QD, as 2D detection histograms (see \supp{2} for QD details). The spectrum over time is seen in (ii), where the time of detection spans the horizontal axis (\SI{10}{ms} time-bins) and energy the vertical axis ($6{-}\SI{7}{meV}$ energy-bins). The effects of thermal broadening, spectral diffusion and blinking, discussed above, can be clearly observed through the width of the spectral peak at each time-bin ($35{-}\SI{50}{meV}$ FWHM), the temporal jitter of the spectral peak position and the variation of emitted intensity, respectively. To achieve spectrally dependent fluorescence decay curves, shown in (iii), the same dataset is analyzed by binning the detections according to their delay from the preceding excitation pulse. Note that a full horizontal binning (FHB) of either histogram is equivalent to a standard spectrometer measurement, and a full vertical binning (FVB) to a standard analysis from a single-SPAD measurement. 

\subsection*{1X spectral dynamics.}
Prior to analyzing the BX state, it is important to first study the 1X state.
Indeed, simultaneous acquisition of both temporal and spectral data enables a more in-depth analysis of the fluorescence spectral dynamics. 
\autoref{fig:specDyn} employs such an analysis to identify and quantify the spectral broadening effects shown in \autoref{fig:intro}b,d for a single QD measurement. \autoref{fig:specDyn}a shows the total fluorescence intensity collected over all array pixels, at \SI{1}{ms} time-bins. The intensity trace features a characteristic blinking behavior - stochastic switching between a bright (`on'), dim (`grey') and dark (`off') fluorescent states \cite{Spinicelli2009,Tenne2013}. The presence of these three states is clearly evident in \autoref{fig:specDyn}b, a histogram of fluorescence intensities over the entire 5-minute measurement. Some time-bins are classified as `mixed', interpreted as time-bins where the QD spent comparable time in different states. Analyzing the spectrum of each intensity state (\autoref{fig:specDyn}c), reveals a clear red-shift of the `grey' state spectra, as well as an order-of-magnitude shorter fluorescence lifetime (see \supp{3}). This supports an identification of the `grey' state as emission from a charged exciton state (identified as a negative trion in past work\cite{Tenne2013}). According to ensemble measurements, the BX state is also expected to present a shorter lifetime and shifted emission compared with the `on' state, making it difficult to differentiate the two contributions.

In addition to blinking, the emission also exhibits spectral diffusion. \autoref{fig:specDyn}d shows the spectral evolution of the neutral 1X emission (`on' state) over time. Each dot represents the momentary mean photon energy over a \SI{1}{ms} time-bin ($\ev{E_{1X}}_{\SI{1}{ms}}$), colored according to the local density of data-points for clarity. The red line represents a Gaussian-weighted ($\sigma=\SI{10}{s}$) moving average of these values ($\ev{E_{1X}}_{\SI{10}{s}}$). This smoothed trend shows a gradual wavelength change, predominantly towards shorter wavelengths, possibly due to oxidation\cite{VanSark2001}. The fast spectral diffusion dynamics are evident in the distribution of $\ev{E_{1X}}_{1ms}$ around this moving average, $\Delta E_{1X}\triangleq \ev{E_{1X}}_{\SI{1}{ms}}-\ev{E_{1X}}_{\SI{10}{s}}$. These faster dynamics are typically attributed to rapid fluctuations in the local electrostatic potential, leading to a shift in emission energy according to the quantum confined Stark effect\cite{Empedocles1997}.

The above results emphasize the difficulty in isolating the BX state emission. Namely, its rather weak contribution is overshadowed by spectral broadening, and especially by the `grey' state emission, which overlaps with it in both spectral and temporal domains. As a result, even a comprehensive analysis of the 2D lifetime-spectrum data (see \supp{4}) was unable to resolve the BX state spectrum.

\subsection*{Heralded spectroscopy.}
To directly probe the BX emission, pairs of photon detections following the same excitation pulse are post-selected. Such paired events are the result of an excitation to the BX state, and two subsequent radiative relaxations (\autoref{fig:intro}a). We note that, due to the low quantum yield of the BX (${\sim}9\%$, see \supp{5} and reference\cite{Lubin2019}), this is not the most probable route for relaxation from the BX state. Yet, its occurrence provides sufficient signal for our analysis. Applying this post-selection (see details in \supp{6}) to the 5-minute single-QD acquisition shown in \autoref{fig:specDyn}, yields ${\sim}\num{1.4d3}$ pairs over \num{1.5d9} excitation pulses (in agreement with theory, see \supp{5}). The 2D spectrum of photon pairs, showing the distribution of the energy of the first emitted photon as a function of that of the second, is shown in \autoref{fig:bx}a. The distribution is clearly centered below the diagonal, indicating BX binding. Note that events where both photons of a single cascade impinge on the same pixel are not detected by the system due to pixel dead time (\SI{{\sim}25}{ns}). \autoref{fig:bx}b highlights the first insight that can be derived by such an approach - the BX spectrum (red dots, FHB of panel a) is red-shifted with respect to the 1X spectrum (blue rings, FVB of panel a). The agreement of the 1X spectrum with the overall spectrum of the `on' state (grey area), corroborates this distinction. The BX binding energy for this particular QD, estimated as the difference between the BX and 1X spectra peaks (extracted by fitting a Cauchy-Lorentz distribution, shown as lines in \autoref{fig:bx}b), is $\varepsilon_b=\num{9.3+-1.0}~\si{meV}$ (68\% confidence interval).

We note that the identification of the 1X and BX spectral peaks is done here without ambiguity. While previous studies required a power dependence series to correctly assign the 1X and BX states\cite{Dekel2000,Oron2006}, heralded spectroscopy obviates this requirement. 
More importantly, this approach super-resolves the few-$\si{meV}$ separated 1X and BX spectral peaks despite their ${\sim}\SI{50}{meV}$ FWHM, and clearly distinguishes between the overlapping BX and `grey' state emission. In fact, owing to the unprecedented sensitivity of this method, measuring QDs featuring lower $\varepsilon_b$ than almost all previous measurements of II-VI semiconductor QDs did not incur any additional challenge (see Discussion section below). 

Thanks to the single-nanocrystal nature of this method, it is not limited to measuring ensemble averaged properties, but can also observe their distribution within the ensemble. \autoref{fig:bx}c shows that the BX binding energy increases with the 1X spectral peak position for 30 QDs taken from the same sample. This can be explained as a result of variation in the physical size of the synthesized QDs. For the QDs investigated in this work, a higher energy 1X spectral peak is likely associated with a thinner CdS shell. A thinner shell also corresponds to further confinement of the electrons in the core and an increased Coulomb interaction between charge carriers, leading to a higher BX binding energy. This trend is in agreement with ensemble measurements for CdSe/CdS seeded nanorods\cite{Sitt2009}.

\subsection*{$\bold{\pmb{\varepsilon}_b-E_{1X}}$ correlation.}
Further insight into the BX state can be obtained from comparing the temporal fluctuations of the 1X and BX spectral peaks. As demonstrated in \autoref{fig:bx}, time-resolved heralded spectroscopy enables isolating the BX energy shift despite the spectral fluctuations. Alternatively, one can refer to the 1X spectral position as a sensor for the micro-environment of the nanocrystal, specifically to the fluctuating local electric field, and observe how the BX binding energy reacts to such fluctuations. \autoref{fig:bx_bySpec}a shows the bivariate distribution of $\varepsilon_b$ and $\Delta E_{1X}$, estimated for each post-selected BX photon event of a single QD `on' state measurement. While the distribution of both variables is widened by the various spectral broadening mechanisms discussed above, one can observe a clear correlation between them. As a guide to the eye, we added red crosses to mark the median binding energy for each \SI{2}{meV} $\Delta E_{1X}$ window. The red line represents a linear fit of these medians, emphasizing the positive cross-correlation of $\varepsilon_b$ and $\Delta E_{1X}$. The slope of this line is $0.59\pm0.08$ (68\% confidence interval); i.e.\ for each \SI{2}{meV} red-shift (blue-shift) of the 1X emission spectral peak, the binding energy is lower (higher) by roughly \SI{1.2}{meV}. \autoref{fig:bx_bySpec}b, a histogram of the $\varepsilon_b$ median slope values for 30 QDs, shows that this positive correlation is evident for all QDs measured. This result suggests that the BX binding energy, much like 1X emission, is subject to the quantum confined Stark effect. A higher local field associated with a red-shift of the 1X emission, is correlated with lower BX binding energy (weaker attraction). This can be attributed to the spatial separation of holes and electrons induced by the external field. Notably, at high enough fields, the sign of the BX binding energy flips, indicating repulsive interaction of the excitons. This observation agrees with past results on the effect of charge separation in type-II QD heterostructures on the BX binding energy\cite{Oron2007}. Furthermore, it strengthens the assertion that spectral diffusion indeed originates from fluctuations in the local electrostatic potential.

\subsection*{Discussion.}\label{sec:discussion}
Due to the practical and fundamental importance of the BX state in II-VI semiconductor nanoparticles, extensive experimental work has been dedicated to the estimation of its energy \cite{Achermann2003,Oron2006,Oron2007,Klimov2007,Sitt2009,Sewall2009,Sewall2011,Geiregat2019,Mueller2021}, with $\varepsilon_b$ values spanning -100's to +10's meV (see \supp{7}). In core/shell CdSe/CdS nanocrystals, $\varepsilon_b$ has been shown to transition continuously from positive (attractive) to negative (repulsive) values when tuning the core or shell diameter\cite{Sitt2009}. This is indicative of a transition from type-I to type-II or quasi type-II architecture, where electrons and holes are separated to different layers of a heterostructure\cite{Oron2007}. The few-meV $\varepsilon_b$ values measured in this work are in good agreement with particles near this transition.

It is worth noting, that for such low $\varepsilon_b$, ensemble measurements become increasingly challenging, as they demand resolving two highly overlapping spectral peaks. Ensemble techniques are, therefore, more readily applied to measure particles exhibiting tens of meV BX binding energies, which are indeed reported more often for II-VI nanocrystals (see \supp{7}). More importantly, ensemble methods require the delicate analysis of a power-series measurement, and therefore prone to systematic biases due to power dependent charging and absorption cross-section heterogeneity. In fact, such heterogeneity in the sample was suggested as the source of recent controversy regarding the BX binding energy in perovskite nanocrystals\cite{Castaneda2016,Shulenberger2019}. The background-free (isolated BX spectrum) and single-particle nature of heralded spectroscopy afford an unprecedented, sub-meV, sensitivity in BX binding energy measurement at room temperature, while at the same time circumventing the above mentioned biases. 

While enabling previously inaccessible measurements, at the present there are two limiting factors that should be taken into account when implementing on-chip SPAD arrays in correlative-spectroscopic setups. First, the detection efficiency (${\sim}10\%$ PDE) is still low compared with common camera alternatives. However, this is in part due to the circuit board design implementation used here, dictating signal collection from only every other pixel and a relatively low gain voltage on the diodes. The next system iteration (currently in development) specifically tailored for the purpose of spectroSPAD, is expected to achieve ${>}30\%$ photon detection efficiency, similar to SPAD arrays implemented with the same technology\cite{Lubin2019}. This will give rise to an order of magnitude increase in the correlation signal level due to its quadratic dependence on the detection probability. Moreover, detection efficiency is complemented by the very low noise level of state-of-the-art SPAD arrays. Unlike commonly used cameras, these arrays avoid readout noise and feature median dark count rates well below 100 counts per second per pixel\cite{Antolovic2018}.
A second factor to consider, common in detector arrays and specific to photon correlation analysis, is inter-pixel crosstalk. Due to the close-packing of pixels, a detection in one pixel has a small probability to lead to a false detection in a neighbouring pixel, and hence a false photon pair. Any bias due to this effect is mitigated here by a combination of the chip design and temporal gating (see \supp{6}), bringing the crosstalk probability down to ${\sim}\num{d-5}$, and of a statistical correction as described in reference\cite{Lubin2019}~.

\section*{Conclusions}
Heralded spectroscopy of BX emission cascades enabled us to perform direct observation and unambiguous identification of emission from multiply-excited states of single QDs at room temperature. In addition to avoiding the pitfalls of indirect and ensemble approaches, by separating the BX and 1X emission in the time domain we greatly extend the range of nanomaterials in which we can observe the BX state, and allow new insights into exciton-exciton interaction within the single nanoparticles. Our study reveals a positive correlation between exciton-exciton attraction and tighter charge-carrier confinement of the single QDs. We also unveil a fluctuation of this attraction strength, correlated with the fast fluctuations of the local electrostatic potential, and significant enough to lead to exciton-exciton repulsion. These capabilities represent a new tool to probe QD physics, and can lead to better design of QD-based technologies where multiexcitonic states typically play a major role.

All this is enabled by constructing the spectroSPAD - a SPAD-based correlative spectrometer extending the temporal resolution limit of standard spectrometers by several orders of magnitude. This tool is not only useful for probing the physics of charge-carrier dynamics, but can also address current challenges in quantum optics and communication.

\subsection*{Supporting Information}
Details of the spectroSPAD system and the linear SPAD array; details of the quantum dots used in this work including synthesis, sample preparation and excitation saturation estimation; fluorescence decay lifetime by intensity state analysis; 2D lifetime-spectrum analysis; BX quantum yield estimation; heralded spectroscopy analysis and correction details; published values of BX binding energies in II-VI nanocrystals.

\begin{acknowledgement}
The authors would like to thank Stella Itzhakov for providing the QDs used in the measurements, Ermanno Giuseppe Bernasconi for characterizing the SPAD array chips and Miri Kazes for assistance with supporting measurements. Financial support by the ERC consolidator grant ColloQuantO 681421, by the Crown center of Photonics and by the Minerva Foundation is gratefully acknowledged. DO is the incumbent of the Harry Weinrebe professorial chair of laser physics.
\end{acknowledgement}

\bibliography{main}

\end{document}

\twocolumn[
\begin{@twocolumnfalse}
\oldmaketitle
\begin{abstract}
This supporting information describes in further detail the data analysis scheme and presents complementary analysis and information to the results described in ``Heralded spectroscopy reveals exciton-exciton correlations in single colloidal quantum dots". The sections are brought in the order of reference in the main text: Details of the spectroSPAD system and the linear SPAD array; details of the quantum dots used in this work including synthesis, sample preparation and excitation saturation estimation; fluorescence decay lifetime by intensity state analysis; 2D lifetime-spectrum analysis; BX quantum yield estimation; heralded spectroscopy analysis and correction details; published values of BX binding energies in II-VI nanocrystals.       
\end{abstract}
\end{@twocolumnfalse}
]

% -----------------------------------------------------------------------------
% ------------------------------- main text -----------------------------------
% -----------------------------------------------------------------------------

\section{spectroSPAD system details}
This section presents a detailed overview of the experimental apparatus, and further technical details of the linear SPAD array detector at its core.

\subsection{spectroSPAD setup overview}
The system is built around a commercial inverted microscope (Eclipse T\textit{i}-U, Nikon). A pulsed diode laser (\SI{470}{nm}, \SI{5}{MHz}, LDH-P-C-470B, PicoQuant) is focused through an oil immersion objective (x100, 1.3 NA, Nikon) on a single QD. Illumination power density at the sample plane is {\SI{{\sim}140}{W/cm^2}} leading to ${\sim}66\%$ probability to excite at least one exciton (per pulse, see \autoref{sec:saturation}). The same objective is used to collect the emitted fluorescence, while back-scattered laser light is filtered by a dichroic mirror (505 LP, Chroma) and a long-pass dielectric filter (488 LP, Semrock). At the output of the microscope, the spectrometer consists of a collimating lens, a blazed grating (235 g/mm, 5.06\textdegree\ blaze, 53-*-790R,  Richardson) and an imaging lens, resulting in \num{3.9d-5} reciprocal linear dispersion and \SI{{\sim}6}{\angstrom} spectral resolution (FWHM). At the spectrometer output image plane, a 512 pixel linear SPAD array (an upgraded version of the sensor described in reference\cite{Burri2017}~, see details in \autoref{sec:linoSPAD}), is placed such that the active pixel pitch is \SI{{\sim}2}{nm} in wavelength (every second pixel is active). An FPGA with an implemented TDC array (synchronized with the laser) assigns timestamps and pixel addresses to single detections in 40 pixels of the array. The trace of detections is then analyzed by a dedicated MATLAB script, implementing temporal and intensity corrections (see \autoref{sec:analysis}) and the analysis schemes.

\subsection{SPAD array technical details}
\label{sec:linoSPAD}

\begin{figure}[h]
    \centering
    \includegraphics[width=.9\linewidth]{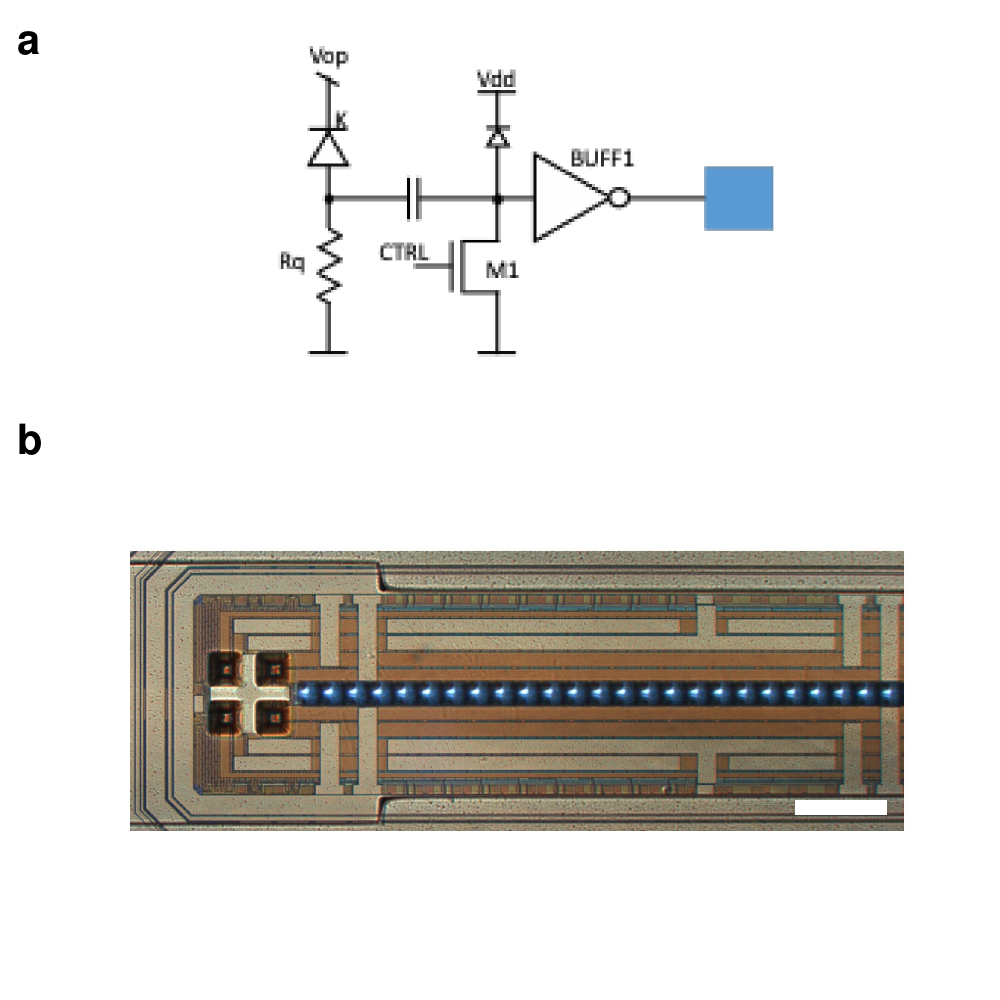}
    \caption{\textbf{The linear SPAD array. a)} Electrical circuit of a SPAD array pixel. \textbf{b)} Optical image of the detector array, mounted with microlenses. Each blue square represents a single pixel. Scale bar is \SI{100}{\mu m}.}
    \label{fig:linoSPAD}
\end{figure}

Since the creation of the first single-photon avalanche diode (SPAD) in complementary metal-oxide semiconductor (CMOS) in 2003, research in the field of SPADs and SPAD image sensors has led to the creation of the first integrated array in 2004, followed by a wide range of SPAD based image sensors with advanced functionality and continuously increasing speed.

The sensor used in this paper comprises an array of 512 SPAD pixels based on the design proposed in reference\cite{Veerappan2016}~. Each pixel comprises a SPAD quenched and recharged passively through a poly resistor. The SPAD is interfaced to the exterior of the chip through a circuit described in \autoref{fig:linoSPAD}a, including capacitive decoupling, a clamp to $V_{dd}$, and a low-threshold buffer. The purpose of this circuitry is to ensure low threshold of detection of the avalanche, thus optimizing jitter while controlling noise. Quenching resistor $R_q$ is designed to present a sufficiently high impedance to the anode of the SPAD, while minimizing the avalanche current, so as to control the overall power consumption of the chip. $V_{op}$ is set to $V_{BD}+V_{EX}$, where $V_{BD}\approx\SI{24.4}{V}$ and $V_{EX}\approx\SI{1.6}{V}$ are the breakdown and excess bias voltages, respectively.

The chip was mounted directly on a board with the SPAD outputs wire-bonded and connected to a field-programmable gate array (FPGA), which hosts the TDCs that enable the time characterization of the response. The TDC array is an improved version of the earlier implementation detailed in reference\cite{Burri2017}~. By multiplexing over 64 TDC channels, up to 256 pixels of the array can be temporally correlated. Extending the system to two FPGA boards, enables simultaneous read out from all 512 pixels. However, to reduce DCR, only 40 pixels were used in this work. The DCR reduction is achieved both by collecting DCR from fewer pixels, and by avoiding the few `hot pixels' in the array that feature exceptionally high DCR (typically `hot pixels' are defined as those that feature DCR at least two orders of magnitude higher than the median; here the `noisiest' pixel DCR was just a factor of 3 above the median (\SI{104}{cps})). The choice of 40 pixels resulted in overall DCR that is about an order of magnitude lower than the detected `on' state emission in intensity measurements. After temporal gating (see \autoref{sec:analysis}) the number of DCR induced pairs was also about an order of magnitude lower than the total number of detected pairs. Pixels exhibit an average jitter of \SI{105}{ps} (FWHM) and a median DCR of \SI{33}{cps}, their native fill factor (without microlenses) and pitch are 25.1\% and 26.2µm, respectively. \autoref{fig:linoSPAD}b shows a detail of the SPAD array, including microlenses. Microlenses were deposited on the chip to enhance effective fill factor and thus overall photon detection efficiency (PDE).

\begin{figure}[h]
    \centering
    \includegraphics[width=.9\linewidth]{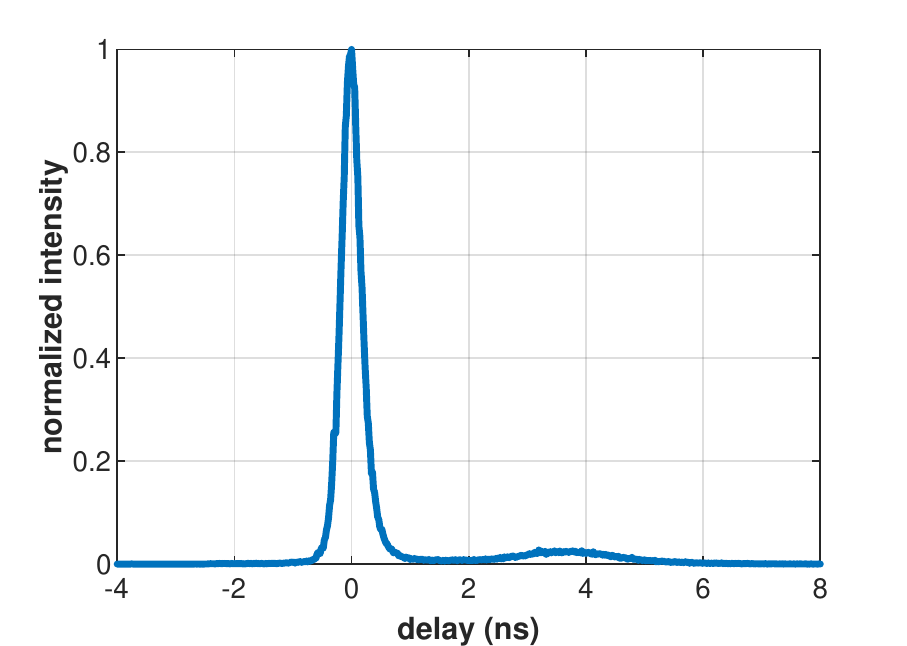}
    \caption{\textbf{Temporal response function of a typical pixel. } A pixel of the array is illuminated directly with the synchronized excitation laser (laser pulses are \SI{<160}{ps} FWHM), and recorded through the TDC. Delay values represent the delay from the response peak.}
    \label{fig:IRF}
\end{figure}

The recorded temporal response after the TDC (\autoref{fig:IRF}) features two peaks: A narrow main peak accounting for ${\sim}90\%$ of the counts, and a secondary broader peak delayed by \SI{{\sim}3.5}{ns}, accounting for the remaining ${\sim}10\%$. The presence of the secondary peak is specific to the particular implementation used in this work, and is not generally evident in similar detector arrays. Preliminary results suggest that this irregular temporal response can be eliminated by better firmware design in the next system iteration (currently in development). The QDs investigated in this work featured 1X decay lifetimes significantly longer than this artifact (see \autoref{sec:specLtFit}). Hence, any ambiguity in order of arrival can be negated by temporal gating (as detailed in \autoref{sec:analysis}) with minimal loss of signal.

\section{Quantum dots used in this work}
\label{sec:QD}
This section describes the synthesis of the QDs used in this work, the sample preparation scheme and an estimation of excitation saturation.

\subsection{Quantum dot synthesis and sample preparation}
\begin{figure}[h]
    \centering
    \includegraphics[width=.8\linewidth]{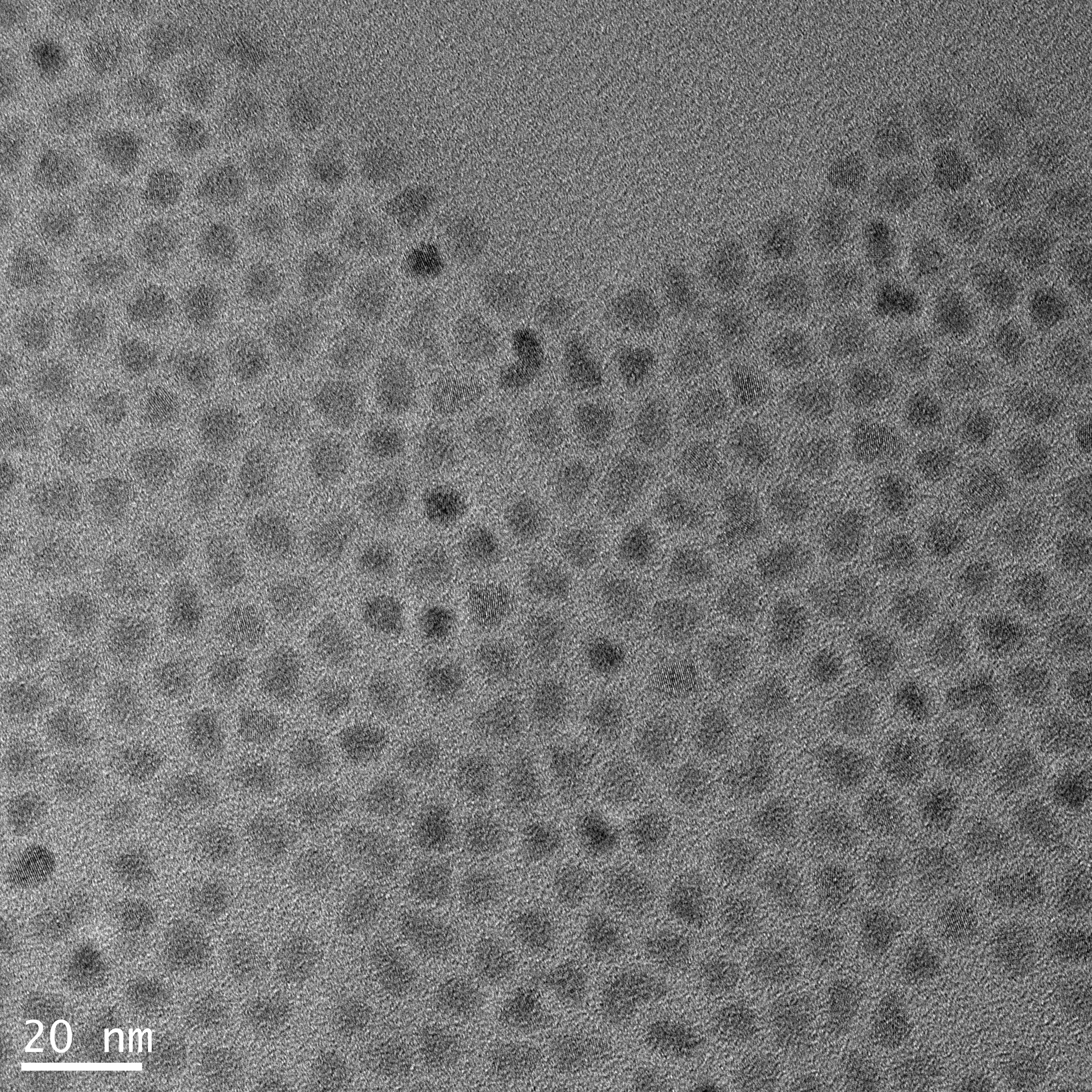}
    \caption{\textbf{Transmission electron micrograph of the QDs used in this work.}}
    \label{fig:tem}
\end{figure}

Colloidal CdSe/CdS/ZnS core/shell/shell QDs were synthesized by the following protocol: A cadmium oxide (CdO), n-tetradecylphosphonic acid (TDPA), and 1-octadecene (ODE) mixture was heated to 280\textdegree C in a three-neck flask under argon environment. Next, a stock solution of trioctylphosphine selenium (TOPSe) was rapidly injected. The temperature was then reduced to 250\textdegree C until the particles reached the desired diameter. A layer-by-layer growth technique in a one-pot synthesis method\cite{Li2003} was used for shell growth of cadmium sulphide (CdS) and zinc sulphide (ZnS). This resulted in spherical QDs with an outer diameter of $5.3\pm\SI{0.6}{nm}$ (see \autoref{fig:tem}). Some of the QDs ($<10\%$) are slightly elongated up to an aspect ratio of 1:1.5. The quantum yield (QY) was measured with an absolute photoluminescence QY spectrometer (Quantaurus-QY, Hamamatsu), and is ~90\%.

Samples were prepared by spin coating a glass coverslip with a solution of these QDs dispersed in a 3wt\% solution of poly(methylmetacrylate) (PMMA) in toluene.

\subsection{Quantum dot excitation saturation}\label{sec:saturation}

\begin{figure}[h]
    \centering
    \includegraphics[width=\linewidth]{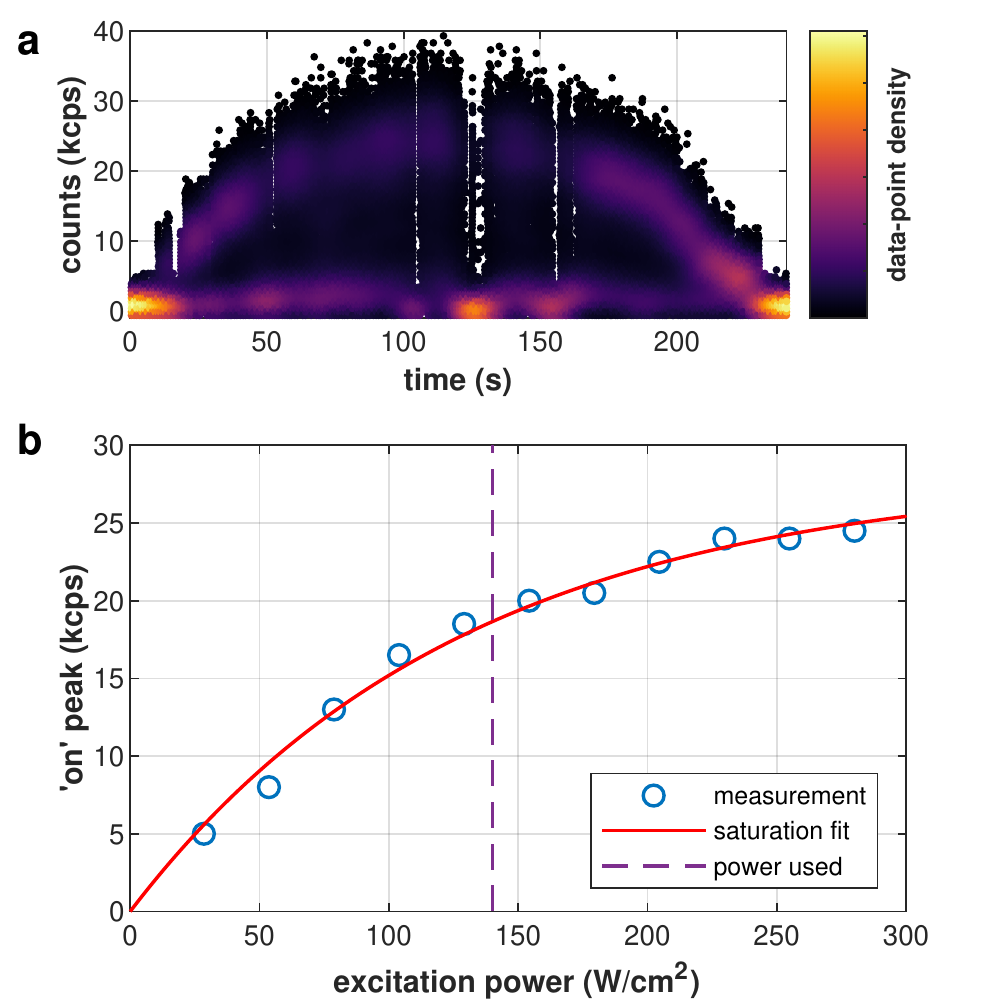}
    \caption{\textbf{Quantum dot excitation saturation. a)} Saturation measurement. A single QD illuminated at increasing intensities from \SI{28}{W/cm^2} to \SI{280}{W/cm^2} and then back down (\SI{10}{s},  \SI{{\sim}25}{W/cm^2} steps). Each point represents the detected intensity at a \SI{5}{ms} time-bin, colored according to the local density of data-points for clarity. \textbf{b)} Peak occuring `on' state intensity for each illumination power (blue circles) and a fit to a saturation curve (red solid line). The excitation power used in this work is marked by a purple dashed line.}
    \label{fig:saturation}
\end{figure}

To estimate QD excitation saturation, single QDs were illuminated at varying intensities. \autoref{fig:saturation}a shows such an experiment, where over time the excitation intensity was increased at a steady rate from \SI{28}{W/cm^2} to \SI{280}{W/cm^2} and then back down. Higher illumination intensities typically result in more time spent in the `grey' and `off' states. Hence, to assess single-excitation saturation, it is essential to identify and estimate the peak occurring intensity of the `on' state alone. This was achieved by creating an intensity histogram at each value of the excitation power, smoothing the histogram with a Gaussian filter and finding the smoothed histogram peak. The peak occurring values are shown in \autoref{fig:saturation}b, together with a fit to a saturation model\cite{Teitelboim2016}:
\begin{equation}
    P = A \cdot \left( 1-exp^{-\frac{I}{I_{sat}}} \right),     
\end{equation}
where $P$ is the `on' state peak, $I$ excitation power, $I_{sat}$ saturation power and $A$ asymptotic `on' state peak (last two are the fit parameters). This simple model assumes negligible contribution to the intensity from multiexcitonic recombination, an assumption justified by the measured value of $g^{(2)}(0)\sim0.1$ (see \autoref{fig:g2}). The data agrees well with the fit and $I_{sat}=129\pm\SI{12}{W/cm^2}$ (68\% confidence interval). The probability to excite $n$ excitons following a single pulse can be estimated from the Poissonian distribution:
\begin{equation}
    (n) = \frac{\lambda^n}{n!}e^{-\lambda}
\end{equation} 
using the distribution parameter $\lambda = \frac{I}{I_{sat}}$. At the excitation power used in this work (\SI{{\sim}140}{W/cm^2}, dashed purple line in \autoref{fig:saturation}), the probability of exciting at least a single exciton following a laser pulse is ${\sim}66\%$ and of exciting at least twice to form a biexciton ${\sim}30\%$.

\section{Lifetime by intensity state}\label{sec:grey}

\begin{figure}[h]
    \centering
    \includegraphics[width=.9\linewidth]{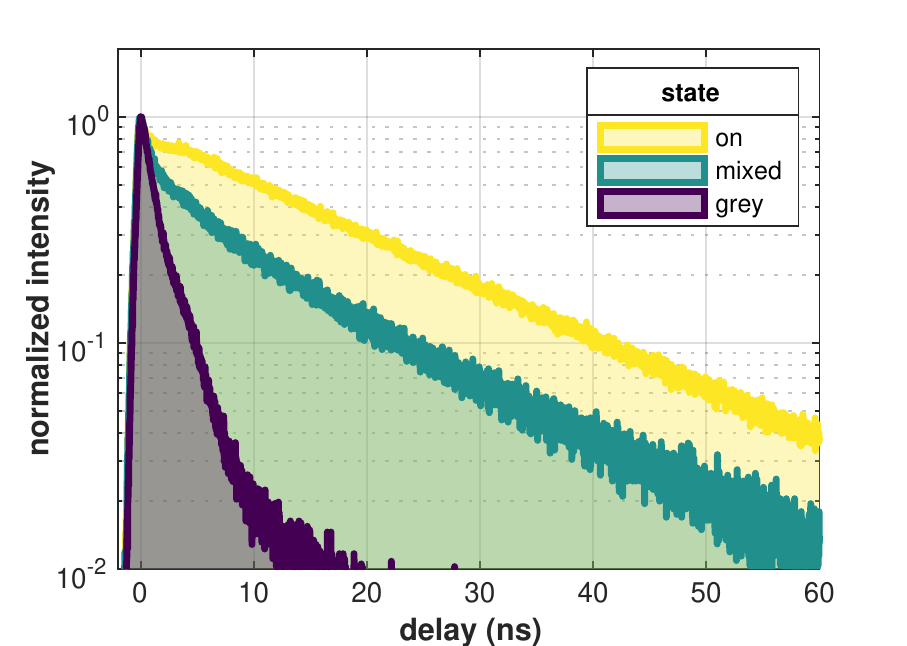}
    \caption{\textbf{Fluorescence decay by intensity state.} Histogram of detection delays from the excitation pulse for the intensity states identified in Figure 2 of the main text. Note the different decay lifetime scales of the `on' and the `grey' state.}
    \label{fig:greyLT}
\end{figure}

\autoref{fig:greyLT} shows fluorescence decay curves for the different intensity states identified in Figure 3 of the main text. Each trace is a sum over all detector pixels for a specific intensity state. The `on' and `grey' state decay traces are both dominated by a single exponential term. However, the `grey' state decay is an order of magnitude faster ($\tau_{on}\approx\SI{20}{ns}$ and $\tau_{grey}\approx\SI{1.35}{ns}$, see fit details in next section). An irregular instrument response function (IRF, see \autoref{sec:linoSPAD}) results in an artifact visible at the first \SI{{\sim}4}{ns} of each curve. 

\section{2D lifetime-spectrum analysis} \label{sec:specLtFit}

\begin{figure}[h]
    \centering
    \includegraphics[width=\linewidth]{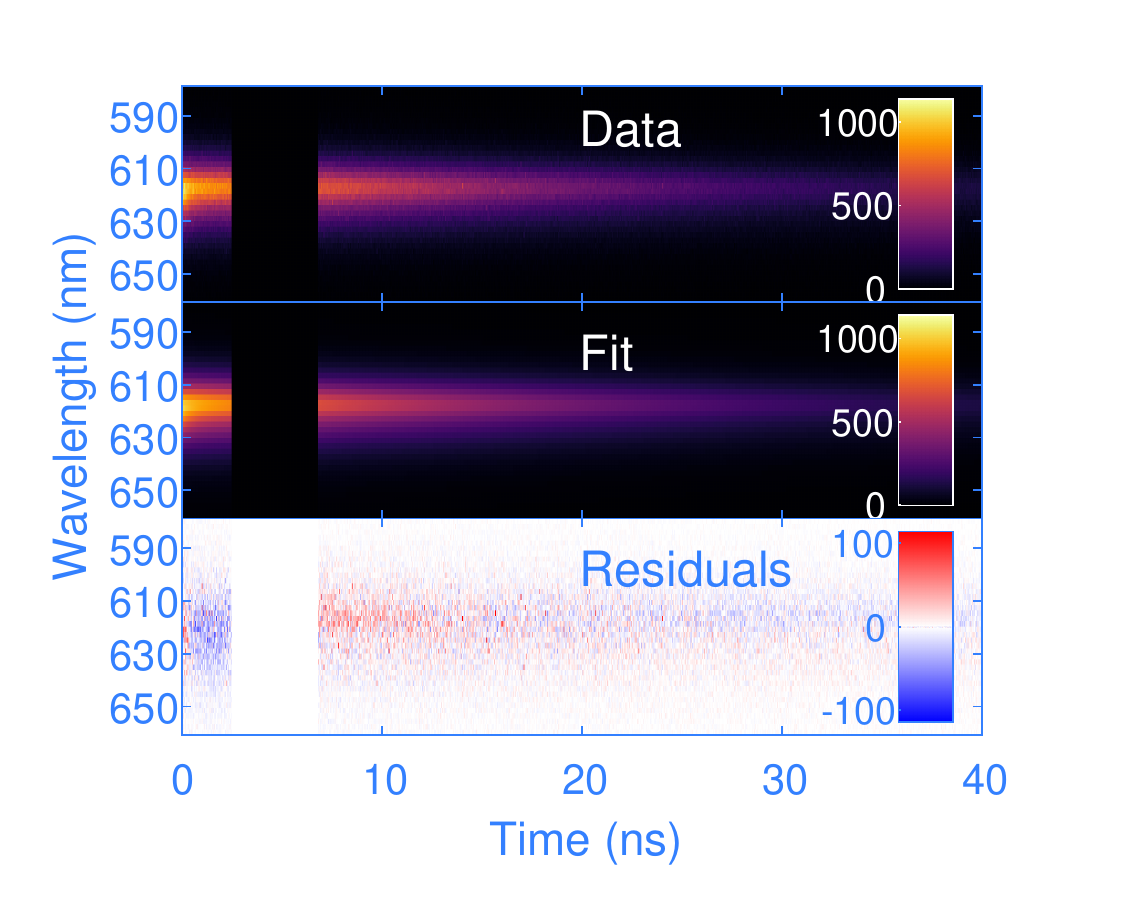}
    \caption{\textbf{2D spectrum-lifetime data and fit.}
    \textbf{Top:} A 2D histogram of photon detections according to their energy (pixel) and arrival time with respect to the excitation laser pulse for the same measurement analyzed in Figures 2-5 of the main text. Here, we include only detections during the `on' intensity state (see Figure~2 in the main text). \textbf{Middle:} A fit of the spectrum-lifetime data. Data in all energies (rows) is fit with the same two exponential terms, allowing the amplitudes to be different for every energy. \textbf{Bottom:} residuals between the experimental data and the fit. The missing band of time delays in all data sets were deducted from the fit in order to mitigate the effect of the irregular IRF (see \autoref{sec:linoSPAD}).}
    \label{fig:specLtFit_2D}
\end{figure}

\begin{figure}[h]
    \centering
    \includegraphics[width=\linewidth]{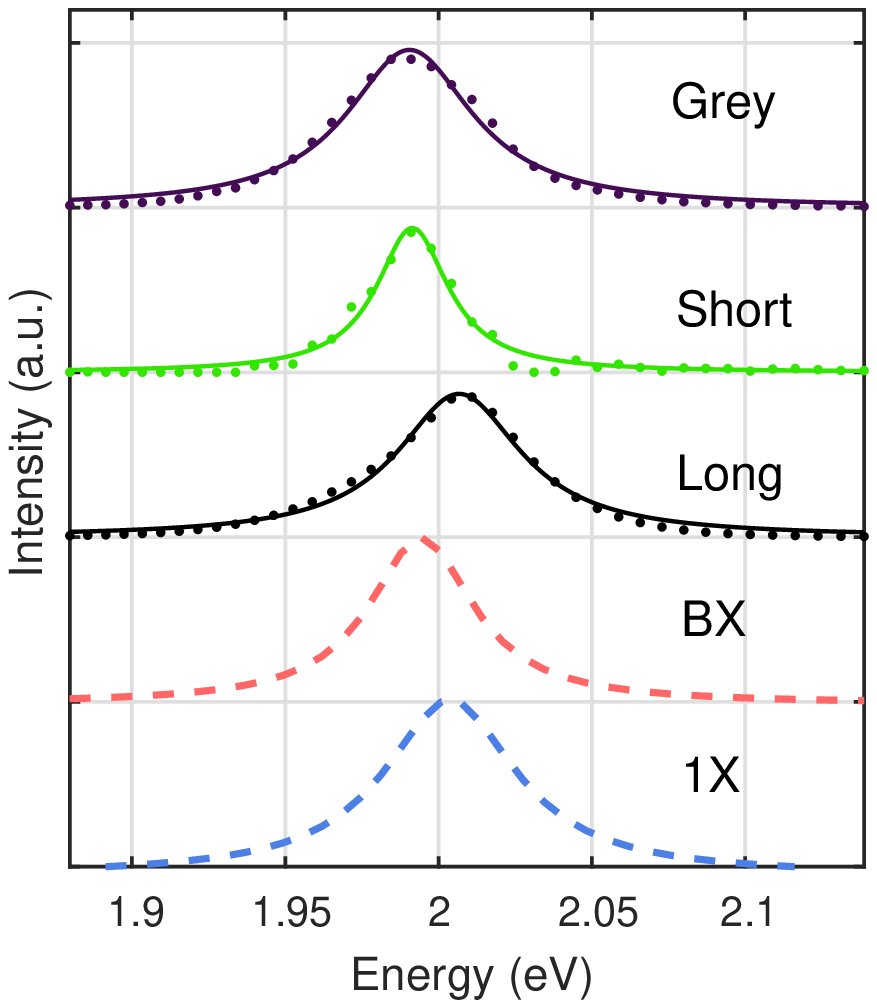}
    \caption{\textbf{The spectra of different lifetime components.}
    Following the fit procedure described in \autoref{sec:specLtFit}, we obtain a spectrum for each lifetime component separately. The black and green dots show the long and short lifetime spectral contributions to the `on' intensity state data, respectively. The purple dots show the spectrum of the charged state obtained through the fit of the `grey' intensity state spectrum-lifetime data. The matching color lines present Cauchy-Lorentz fits for each spectrum. For comparison with the heralded spectroscopy method, described in the main text, we present the fit of the 1X (BX) spectrum for the same QD with a blue (red) dashed line (as shown in Figure 4b of the main text}
    \label{fig:specLtFit_spectra}
\end{figure}
Apart from heralded spectroscopy, presented in the main text, the single-particle, spectro-temporal information provided by the spectroSPAD can reveal connections between the spectral and the dynamical characteristics of nanocrystal fluorescence. One example of such an observation is presented in the 2D histogram shown in the top panel of \autoref{fig:specLtFit_2D}: Photon detections (here only from the 'on' blinking state), are binned according to both their arrival time (with respect to excitation pulse) and their energy. In such a spectrum-lifetime dataset, one can differentiate the spectra of different lifetime components. While this type of data is commonly measured for an ensemble of particles (with a scanning monochromator), this is the first demonstration of such a measurement for a single QD. 

In the following we attempt to distill the spectrum and lifetime of the biexciton (BX) state from the multi-component data. We show that even when observing the `on' state data alone, it is extremely difficult to separate the contributions of the charged and the BX state. As a result, we claim that determining the BX binding energy from such an analysis is a challenging task, and is prone to ambiguities.

To analyze the spectrum-lifetime data, we perform a global fit with the same two exponential decaying components for each of the energy-bins (detector pixels). That is, we fit every row in the matrix presented in the top panel of \autoref{fig:specLtFit_2D}, with a sum of two decaying exponential functions. As a result, for $n_{pix}$ pixels, we have $2+2{\cdot}n_{pix}$ fit parameters, one lifetime parameter and $n_{pix}$ amplitudes (denoting the spectrum) for each fit component. Note, that in order to reduce the effect of the irregular IRF (see \autoref{sec:linoSPAD}), we omit a $4.3$ ns portion of the data in all fits (the apparent time gap in \autoref{fig:specLtFit_2D}).

Fitting the `grey' state spectrum-lifetime data (not shown here), we observe that most of the contribution ($>80\%$) arises from a short lifetime component (\SI{1.35}{\ns}) which we associate with the well-known short lifetime of the charged state. The respective spectrum of this state is shown in \autoref{fig:specLtFit_spectra} (purple dots) together with a fit to a Cauchy-Lorentz distribution centered at \SI{1.99}{\eV} (purple line).
Next, we turn our focus back to the `on' state spectrum-lifetime data shown in the top panel of \autoref{fig:specLtFit_2D}. Fitting the data with two decaying exponentials, we obtain a short (${\sim}\SI{0.5}{\ns}$) and a long (${\sim}\SI{20}{\ns}$) lifetime components. The long lifetime matches the common lifetime of the single exciton state (1X) in CdSe QDs and indeed its associated spectrum (\autoref{fig:specLtFit_spectra}, black dots) resembles that of the `on' state spectrum presented in figures 3 and 4 of the main text. The peak of a fitted Cauchy-Lorentz distribution (black line) is at $2.007$ eV, very close to $2.004$ eV obtained from the fit of the heralded `on' state spectrum.

Unlike the straightforward analysis of the long lifetime contribution, interpreting the short lifetime component is quite a challenging task. First, its integrated intensity is very small, ${\sim}1/400$ of that of the 1X term. In addition, at short time delays the residuals of the fit (\autoref{fig:specLtFit_2D}, bottom panel), are substantial and systematic (switch from positive to negative sign around $0.4$ ns), indicating that our model does not fully account for all features. Most importantly, the spectrum of the short lifetime component (\autoref{fig:specLtFit_spectra}, green dots) coincides with that of the charged state.

Considering the intensity proportion of the short lifetime component, the expected contribution to the data from BX recombination is given by 
\begin{equation}
    \frac{N_{2ph}}{N_{1ph}} =
    \frac{p(N\geq 2)}{p(N\geq 1)}\cdot g^{(2)}(0) \approx \\ 0.05, 
\end{equation}
where $g^{(2)}(0)$ is the second-order correlation function at zero delay time (antibunching factor) and $p(N\geq k)$ is the probability to populate $k$ or more excitons after a single laser pulse, given by a Poisson distribution. In the case of the present measurement, the probability of exciting at least a single exciton was $66\%$ and that of at least two excitons was $30\%$ (see \autoref{sec:saturation}). The expected calculated intensity proportion is more than an order of magnitude larger than that of the short lifetime fit component.

It is therefore evident that the short lifetime fit component in the 'on' state doesn't match the expected contribution of the BX in both the spectral position and in amplitude. We could not pinpoint the reason for the lack of a fast decaying signal for the BX recombination in the spectrum-lifetime fit. However, we suspect that both the short lifetime of the BX decay and the systematic errors in the fit result in a lack of accuracy for such an analysis. For example, even slight differences in IRF between different pixels can result in a systematic bias that can be overcome only with a complex characterization and modeling procedure. 

While a more sophisticated numerical analysis protocol (e.g.\ taking the IRF into account) may lead to better results, the above analysis shows the challenges of interpreting such a data set. Namely, the contribution of the BX state is extremely weak and its characteristics overlap other states. For example, a fit that includes three decaying exponentials for the `on' state resulted in over-fitting of the results and the emergence of unnatural spectral features. In contrast, heralded spectroscopy, presented in Figure 4 of the main text, unambiguously isolates the BX component from the 1X and charged `grey' state, enabling a reliable and robust estimation of the BX spectra.

\section{Biexciton quantum yield}
\begin{figure}[h]
    \centering
    \includegraphics[width=.9\linewidth]{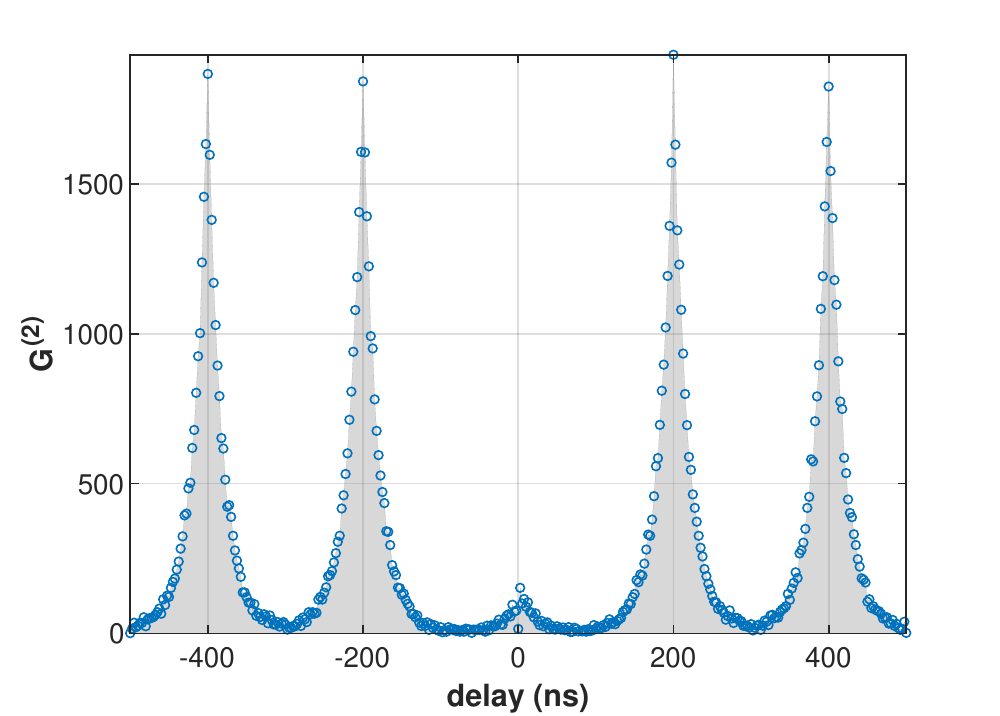}
    \caption{{\bf Second order photon correlations.} Second order photon arrival time correlations extracted from the measurement featured in Figures 3-5 of the main text. The histogram uses \SI{1.25}{ns} delay bins.}
    \label{fig:g2}
\end{figure}
The BX QY to 1X QY ratio can be estimated from the second order correlation of photon arrival times seen in \autoref{fig:g2}. The figure was generated and corrected by applying the method described in reference\cite{Lubin2019} to the spectroSPAD measurements. Briefly, the measurement described in the main text can be viewed as a multiple arm Hanbury Brown and Twiss correlation setup, where the combination of spectral thermal broadening and the spectrometer are used instead of beamsplitters to multiplex the detection. Second order correlation of arrival times is calculated by histograming delays between detections (arrival times) in different pixels, and temporal and intensity corrections are calculated from the same data used in \autoref{sec:analysis}. The result are peaks separated by the pulse repetition period (\SI{200}{ns}), and broadened by the fluorescence decay lifetime ($\tau{\sim}\SI{20}{ns}$). The zero delay peak is significantly lower indicating antibunching - the lower probability of two photon detections following the same pulse. The ratio of BX QY and 1X QY, is equal to the normalized second order correlation at zero time delay ($g^{(2)}(0)$), calculated as the ratio between the zero delay peak area and the mean area of all the non-zero delay peaks. The value for this specific QD estimated from \autoref{fig:g2} is $g^{(2)}(0)=0.09\pm0.02$ (standard error). The artifact seen near zero time delay originates from shot noise on the significant crosstalk correction. A thorough study of identical QDs from the same synthesis with a more accurate setup showed that this value is within the expected range of $g^{(2)}(0)=0.10\pm0.02$ (standard deviation)\cite{Lubin2019}. The 1X QY is ${\sim}90\%$ (see \autoref{sec:QD}), and hence the BX QY is ${\sim}9\%$.

The ratio between heralded photon pair detections and all detections ($\alpha_{BX}$), under the excitation conditions used in this work (see \autoref{sec:saturation}), can be estimated as
\begin{multline}
    \alpha_{BX} \sim
    \frac{N_{2ph}}{N_{1ph}} \sim
    0.45\frac{g^{(2)}(0) \cdot 0.5 \cdot p_{det}^2}{p_{det}} \approx \\ 0.23 \cdot p_{det}\cdot g^{(2)}(0), 
\end{multline}
where $p_{det}$ is the probability to detect a photon per laser pulse from a single QD. The factor of $0.45$ is the result of the ratio between of the probability of exciting at least a single exciton ($66\%$) and that of at least two excitons ($30\%$) (see \autoref{sec:saturation}). During the `on' state of the QDs measured in this work, values of $p_{det} \sim 10^{-2}$ and $g^{(2)}(0)\sim 10^{-1}$ are measured, resulting in  $\alpha_{BX}\sim 2\cdot 10^{-4}$. This value is in excellent agreement with the $\alpha_{BX}$ value extracted by the heralded approach ($(2.0\pm0.7) \cdot 10^{-4}$ for the 30 QDs shown in Figures 4c and 5b of the main text), further corroborating the reliability of these results.

\section{Analysis details}\label{sec:analysis} 
This section details the heralded spectroscopy parameters and the temporal and intensity corrections appended to the acquired data.

\subsection{Heralded spectroscopy} Photon pairs were time gated to support correct identification of BX and 1X emission and reduce the contribution of background due to dark counts. BX emission was gated to the first \SI{5}{ns} following the excitation pulse. Due to the short lifetime of the BX state, this gating leads to a negligible loss of signal accompanied with a significant reduction of noise. 1X detections were gated to $5-\SI{60}{ns}$ delay from the BX detection. The upper bound serves to reduce noise while accommodating the longer lifetime of the 1X state emission (see \autoref{sec:specLtFit}). The lower limit filters out possible misidentification of BX and 1X due to the IRF (see \autoref{sec:linoSPAD}). The upper limits for BX and 1X also assert that only photon pairs following the same pulse are taken into account. Emission spectral peaks were then estimated by a fit to a Cauchy–Lorentz distribution.

In Figure 4 of the main text, $E_{BX}$ and $E_{1X}$ correspond to the energies of the first and second photons, respectively, of the pairs identified by the time gating described above. In Figure 5a of the main text, the momentary mean 1X energy ($\ev{E_{1X}}_{\SI{1}{ms}}$) and averaged 1X energy ($\ev{E_{1X}}_{\SI{10}{s}}$) are estimated at the time of each such pair event (as described in Figure 3d of the main text). $\Delta E_{1X}\triangleq \expval{E_{1X}}_{\SI{1}{ms}}-\expval{E_{1X}}_{\SI{10}{s}}$ is used as an estimator of the momentary spectral shift (horizontal axis). The momentary BX binding energy is estimated as the difference between $\ev{E_{1X}}_{\SI{1}{ms}}$ and $E_{BX}$ (vertical axis). Using these estimators (i.e.\ $\ev{E_{1X}}_{\SI{1}{ms}}$ instead of $E_{1X}$) is beneficial when estimating momentary spectral shift and binding energy for each pair event separately, as the \SI{1}{ms} averaging averages over the thermal broadening distribution of 1X emission, reducing its effect on the final results.

\subsection{Temporal Corrections} The time-to-digital converter (TDC) architecture assigns timestamps with a mean interval of \SI{{\sim}18}{ps} (the detector jitter is larger, see \autoref{sec:linoSPAD}). However, as detailed elsewhere\cite{Burri2017}, the timestamps are not uniformly spaced but rather each span a $0-\SI{92}{ps}$ range of arrival times (most spans are within $18\pm\SI{12}{ps}$). This non-uniformity was characterized by illuminating the detector with temporally featureless halogen light, and recording the occurrence of each timestamp as a measure of the relative time duration it spans. The correction was then implemented statistically by assigning to each recorded raw-timestamp a corrected-timestamp chosen at random from the respective time span. In addition, timestamps recorded for each detector pixel are differently delayed from the TDC trigger. This was characterized by illuminating the detector directly with the \SI{<160}{ps} FWHM excitation laser pulse, and adding a per-pixel timestamp delay so that the recorded pulse peaks in all detectors temporally overlap.

\subsection{Intensity corrections} Two sources of false detections and detection pairs had to be considered in the analysis of heralded spectroscopy. The first, dark count rate (DCR), was recorded and subtracted from the intensity trace (per pixel). The expected number of DCR-photon detection pairs was estimated and subtracted from the photon pair signal (DCR-DCR pair occurrence is negligible). The second source of false photon pairs, detector crosstalk, was characterized and corrected statistically, by the protocol detailed in reference\cite{Lubin2019}.

In Figure 5a of the main text, each data-point represents a single photon pair event. Hence, only in this figure, intensity corrections could not be implemented with this statistical approach. The number of DCR and crosstalk induced pairs in this figure can be estimated to be ${\sim}10\%$ of the overall data-points. Hence, the use of a median estimator (red crosses in Figure 5a of the main text) mitigates the effect of noise induced outliers. Furthermore, to avoid biases where noise might be more significant than signal, the median values shown and considered for the fit were only for $\Delta E_{1X}$ energy-bins including at least 1\% of the total signal.

All corrections were verified to be stable over time.

\section{Published values of BX binding energy}

\begin{table*}[h]
\centering
    \begin{tabular}{M{3.4cm} M{2cm} M{3cm} M{2cm} M{1cm} M{2.25cm} M{0.75cm}}
    \toprule
         Material (core/shell) &Shape &Size (radius in nm) &Method*  &Year  &BX binding energy (meV)  &Ref  \\
         \midrule       
         CdSe/ZnS &dots &1-3.5 core &uPL &2003 &\textbf{10 - 30} &\citenum{Achermann2003} \\
         
         CdSe &dots &2.4-4 &uPL &2005 &\textbf{28 - 38} & \citenum{Bonati2005}\\
         
         CdSe/ZnS &dots &1.25-4.5 core &TRPL &2006 &\textbf{20 - 50} &\citenum{Oron2006} \\
         
         CdTe/CdSe and CdSe &dots &1.9/0-2.5 core/shell &TRPL &2007 &\textbf{-30 - 30} & \citenum{Oron2007}\\
         
         CdS/ZnSe &dots &1.6/2 core/shell &TRPL &2007 &\textbf{-106} & \citenum{Klimov2007}\\
                 
         CdSe &dots &1.8-3 &TA &2008 &\textbf{18 - 25} &\citenum{Sewall2008} \\
         
         CdSe:Te doped &dots &1-2 &TRPL &2008 &\textbf{-300 - -100} &\citenum{Avidan2008} \\
                  
         CdSe &dots &1.5-2.8 &TA &2009 &\textbf{9 - 18} &\citenum{Sewall2009} \\
                  
         CdSe &dots &1.5-2.8 &uPL &2009 &\textbf{37 - 50} &\citenum{Sewall2009} \\
                  
         CdSe/CdS &rods &1.1-2/(2x50) &TRPL &2009 &\textbf{-40 - 30} &\citenum{Sitt2009} \\
         
         CdS/ZnSe &dots &1.9 &TRPL &2010 &\textbf{-70 - -20} & \citenum{Ivanov2010}\\
                  
         CdSe/CdS &giant dots &3/8 &CS &2010 &\textbf{-25} & \citenum{Htoon2010}\\
                  
         CdSe &dots &2.8 &TA &2011 &\textbf{10} &\citenum{Sewall2011} \\
         
         CdSe/ZnS &rods &- &CS &2011 &\textbf{12} &\citenum{Louyer2011} \\
         
         CdSe &nanoplatelets &0.67x5x17 &TGPL &2019 &\textbf{45} &\citenum{Geiregat2019} \\
         
         CdSeS/ZnS &dots &3 core+shell &2DFS &2021 &\textbf{16} & \citenum{Mueller2021}\\
         
         CdSe &nanoplatelets & 0.6x3.5x10.5 &CS &2021 &\textbf{16.5} &\citenum{Peng2021}\\
         \midrule
         CdSe/CdS/ZnS &dots &5.3 core+shell &HS &2021 &\textbf{6} & This work\\         
         \bottomrule
    \end{tabular}
    \caption{Published values for BX binding energy in II-VI nanocrystals}
    \label{tab:LitBX}
    \footnotesize{* \textbf{uPL} - PL upconversion, \textbf{TRPL} - Time resolved PL, \textbf{TA} - Transient absorption, \textbf{CS} - Single particle PL spectroscopy at cryogenic temperatures, \textbf{TGPL} - Transient grating PL, \textbf{2DFS} - Two dimensional fluorescence spectroscopy, \textbf{HS} - Heralded spectroscopy.\vspace{110pt}}
\end{table*}

In order to compare the measured value of BX binding energy in this work to previously reported values in II-VI nanoparticles we have summarized a survey of the literature in \autoref{tab:LitBX}. We note that different works use opposite conventions with respect to the definition of sign in biexciton binding energy. Here, we adopt the convention in which a positive binding energy ($\varepsilon_b>0$) refers to a BX spectral peak at a lower energy with respect to the 1X spectral peak due to attractive interaction between the excitons. In the case of single particle techniques, such as the current work, we note only the ensemble average BX binding energy in the table.

\cleardoublepage
\bibliography{supp}